\documentclass[twocolumn]{aastex62}
\usepackage{amsmath}

\include{definitions} 
\received{\today}

\shorttitle{MASS RATIO OF MAXI~J1820+070}
\shortauthors{Torres, M. A. P. et al. }

\begin{document}

\title{THE BINARY MASS RATIO IN THE BLACK HOLE TRANSIENT MAXI~J1820+070}


\author[0000-0002-5297-2683]{M. A. P. Torres}
\affiliation{Instituto de Astrof\'isica de Canarias, E-38205 La Laguna, Tenerife, Spain}
\affiliation{Departamento de Astrof\'\i{}sica, Universidad de La Laguna, E-38206 La Laguna, Tenerife, Spain}
\author[0000-0001-5031-0128]{J.~Casares}
\affiliation{Instituto de Astrof\'isica de Canarias, E-38205 La Laguna, Tenerife, Spain}
\affiliation{Departamento de Astrof\'\i{}sica, Universidad de La Laguna, E-38206 La Laguna, Tenerife, Spain}
\author{F. Jim\'enez-Ibarra}
\affiliation{Instituto de Astrof\'isica de Canarias, E-38205 La Laguna, Tenerife, Spain}
\affiliation{Departamento de Astrof\'\i{}sica, Universidad de La Laguna, E-38206 La Laguna, Tenerife, Spain}
\author[0000-0002-0621-1293]{A. \'Alvarez-Hern\'andez}
\affiliation{Instituto de Astrof\'isica de Canarias, E-38205 La Laguna, Tenerife, Spain}
\affiliation{Departamento de Astrof\'\i{}sica, Universidad de La Laguna, E-38206 La Laguna, Tenerife, Spain}
\author[0000-0002-3348-4035]{T.~Mu\~noz-Darias}
\affiliation{Instituto de Astrof\'isica de Canarias, E-38205 La Laguna, Tenerife, Spain}
\affiliation{Departamento de Astrof\'\i{}sica, Universidad de La Laguna, E-38206 La Laguna, Tenerife, Spain}
\author{M. Armas Padilla}
\affiliation{Instituto de Astrof\'isica de Canarias, E-38205 La Laguna, Tenerife, Spain}
\affiliation{Departamento de Astrof\'\i{}sica, Universidad de La Laguna, E-38206 La Laguna, Tenerife, Spain}
\author[0000-0001-5679-0695]{P.G.~Jonker}
\affiliation{SRON, Netherlands Institute for Space Research, Sorbonnelaan 2, NL-3584 CA Utrecht, the Netherlands}
\affiliation{Department of Astrophysics/IMAPP, Radboud University, P.O.~Box 9010, 6500 GL Nijmegen, The Netherlands}

\author[0000-0002-1082-7496]{M.~Heida}
\affiliation{ESO, Karl-Schwarzschild-Str 2, 85748 Garching bei München, Germany}

\begin{abstract}
We present intermediate resolution spectroscopy of the optical
counterpart to the black hole X-ray transient MAXI J1820+070
(=ASASSN-18ey) obtained with the OSIRIS spectrograph on the 10.4-m
Gran Telescopio Canarias. The observations were performed  with the
source close to the quiescent state and before the  onset of renewed activity in August 2019. We make use of these data and K-type dwarf templates taken with the
same instrumental configuration to measure the projected rotational velocity of the donor star. We find $v_{rot} \sin i = 84 \pm 5$ km s$^{-1}$
($1\!-\!\sigma$), which implies a donor to black-hole mass ratio $q =
{M_2}/{M_1} = 0.072 \pm 0.012$  for the case of a tidally locked and
Roche-lobe filling donor star. The derived dynamical masses for the
stellar components are $M_1 = (5.95 \pm 0.22)\sin ^{-3}i$ $M_\sun$ and
$M_2 = (0.43 \pm 0.08) \sin^{-3}i$ $M_\sun$.  The use of $q$, combined with estimates
of the accretion disk size at the time of the  optical spectroscopy, 
allows us to revise our previous orbital inclination constraints to $66^{\circ} < i
< 81^{\circ}$. These values lead to 95\% confidence level limits on
the masses of $5.73 <M_1(M_\odot) < 8.34$ and $0.28 < M_2(M_\odot) <
0.77$.  Adopting instead the $63 \pm 3^{\circ}$ orientation
angle of the radio jet as the binary inclination leads
to $M_1 = 8.48^{+0.79}_{-0.72} M_\odot$ 
and $M_2 = 0.61^{+0.13}_{-0.12} M_\odot$ ($1\!-\!\sigma$).
\end{abstract}


\section{Introduction}\label{sec:intro}

MAXI J1820+070 (hereafter J1820) is an X-ray transient
discovered by the MAXI mission at the rise of its March 2018 outburst
(\citealt{2018ATel11399....1K}).  The source  was soon after 
classified as a transient black hole (BH) candidate in the low-hard state in light of the
multi-wavelength and variability
properties \citep{2018ATel11399....1K, 2018ATel11403....1K,
  2018ATel11418....1B, 2019ApJ...874..183S}.  The source later
  moved to the soft state (e.g. \citealt{2020MNRAS.493.5389F}). During
  the state transition, a change on the type of X-ray quasi-periodic
  oscillations was observed and immediately followed by the launch of
  a relativistic jet \citep{2020ApJ...891L..29H,2020NatAs.tmp....2B}. 

The spectrum of the optical counterpart (ASASSN-18ey; \citealt{2018ATel11400....1D},
\citealt{2018ApJ...867L...9T}) displayed emission lines typical of
low-mass X-ray binaries in outburst with marked disk wind profile components 
\citep{2019ApJ...879L...4M}.  An optical modulation with a period of
$16.87 \pm 0.07$ h was found from intense photometric monitoring
during the early outburst phase, and  interpreted as a superhump (\citealt{2018ATel11756....1P}).  Multi-band counterparts 
were identified in pre-discovery images of the field
\citep{2018ATel11400....1D,2018ATel11418....1B,2018ApJ...867L...9T}. 
The  optical counterpart is sufficiently bright to deliver a parallax
determination in GAIA DR2 (e.g. \citealt{2019MNRAS.485.2642G,
2019MNRAS.489.3116A}) and  historic  outburst episodes have been detected
through  analysis of photographic plates taken in 1898 and 1934
\citep{2019ATel13066....1K}.  Radio astrometric observations of the
Gigahertz counterpart has provided a trigonometric parallax that yields a
direct distance measurement towards J1820  of $2.96 \pm 0.33$ kpc and an inclination to the line of sight for the jet ejecta of  $63 \pm 3^{\circ}$ \citep{2020MNRAS.493L..81A}.

The radial velocity curve of the donor star in J1820 was derived from data
acquired during episodes of fading activity that took place in February and June 2019,  when the optical counterpart dimmed close to 
the pre-outburst brightness \citep{2019ApJ...882L..21T}.  The measured
mass of the compact object exceeds 5.2 M$_\odot$, placing J1820  among
the dynamically established stellar-mass BHs, with only less than 20
known in the Galaxy \citep{2014SSRv..183..223C,2016A&A...587A..61C}.
The spectroscopic orbital period found from the radial velocities was a few percent shorter than the photometric period
reported by  \cite{2018ATel11756....1P}, %
confirming the occurrence of superhumps during outburst. The prompt
determination of the orbital ephemeris also permitted to conclude that
at least one X-ray dipping episode observed during outburst
(\citealt{2019MNRAS.488L..18K}, see also\citealt{2018ATel11576....1H}) took place at an orbital phase  where absorption by the outer disk bulge is expected. \citet{2019ApJ...882L..21T} set the first limits on the binary
orbital inclination to $69^{\circ} \lesssim i \lesssim
77^{\circ}$. These constraints were obtained  by considering the
absence of X-ray eclipses during outburst, the detection 
of a disk grazing eclipse by the donor star in the H$\alpha$ line %
and a provisional value for the mass ratio of $q\simeq
0.12$. The latter was 
an estimate obtained from the empirical relationship between $q$
and the orbital period to superhump period excess found for cataclysmic
variables.  

In this manuscript we present new optical spectroscopy of J1820 that 
leads to a direct and accurate determination %
of $q$. The performed analysis and results are outlined as follows: Section
\ref{sec:obs} describes the observations and the data reduction steps. In Section \ref{sec:q} we measure the projected rotational broadening of the photospheric lines from
the donor star and  $q$. The value of $q$ is utilized in Section
\ref{sec:inc} to set constraints on the accretion disk 
outer radius  and revise the limits for the binary system inclination.  Finally, in Section \ref{sec:disc} we discuss our results.

\section{Observations and data reduction} \label{sec:obs}

The data presented in this paper were obtained with the 10.4-m Gran Telescopio
Canarias (GTC) at the Observatorio del Roque de los Muchachos on La Palma, Spain.
The observations were performed with the OSIRIS spectrograph
\citep{2000SPIE.4008..623C} across 3 different nights in July and
August 2019, when  J1820 was near the quiescent state.  In Table
\ref{log} we provide the seeing conditions and the
r-band apparent magnitude of the optical counterpart. The former was measured from the  spatial profile of the spectra
and  the latter from the OSIRIS acquisition images.
The table also gives  the orbital phases at the time of the observations.

We used grism R2500R to cover the $5575-7685$ \AA~wavelength range  with a 0.5
\AA~pix$^{-1}$ dispersion (unbinned detector).  All observations were carried out with a
0\arcsec\!.4-wide slit in order to secure a slit-limited resolution
of 2.0 \AA~FWHM  (90 km~s$^{-1}$ at H$\alpha$).
In total we were able to collect nine 900~s
spectra of J1820 before it resumed outburst activity.  In order to
evaluate the line rotational broadening from these data, high
signal-to-noise ratio exposures of stars HD 219134 (K3 V), HD 216803 (K4 V) and 61 Cyg A (K5 V) were performed under
slit-limited conditions using the same instrumental setup as for
J1820.  Finally, HgAr+Ne comparison arc lamp spectra were taken at  the end of each 
observing night.

\begin{table}
\begin{center}
\caption{Journal of the J1820 OSIRIS observations.}
\begin{tabular}{lcccc}
\tableline\tableline
Date               & Orbital\footnote[1]{according to the ephemeris in \citep{2019ApJ...882L..21T}.}         & Exp.  & seeing\footnote[2]{FWHM
                                               of the 
                                               spatial profile at spectral positions
                                               near
                                               $\lambda6410$.}
  &   r  \\
(2019)            & phase     &   (\#)      &
                                                                   ('')
                                                                         &
                                                                           (mag\footnote[3]{photometry calibrated
with field stars in Pan-STARRS \citep{2016arXiv161205560C}.}) \\
\\
29  Jul & 0.28-0.32     & $3$ & 0.7-1.2 & 17.8-17.5 \\
3 Aug &  0.50-0.53    & $3$ & 0.8-1.0 & 17.6-17.4 \\
7 Aug &  0.36-0.39   &  $3$ & 0.8-0.9 & 17.4 \\

\tableline
\end{tabular}
\end{center}
\label{log}
\end{table}

The spectra were reduced, extracted and wavelength calibrated following
standard techniques implemented in {\sc  iraf}. The pixel-to-wavelength
scale was determined through third order spline fits to 12 arc lines. The
rms scatter of the fits were always $<0.01$ \AA. We made use of the [O{\sc i}]  6300.3
\AA~ sky emission line to  correct for wavelength zeropoint
deviations, which were $<19$ km s$^{-1}$. 

\section{Analysis and results}\label{sec:res}

To perform the analysis detailed in this section we made use of {\sc molly} and {\sc
  python}-based software.  Unless otherwise stated, to evaluate the 1-$\sigma$
uncertainty in $q$ and other quantities we applied a Monte Carlo
approach where we draw 10,000 random values from normal distributions
defined  by the mean value and  1-$\sigma$ errors of the measurements in play. 
We also employed  the ephemeris and orbital
elements given by \citet{2019ApJ...882L..21T} since they were not significantly  improved by the inclusion of radial velocities derived from
the  data under discussion. Finally, for the analytical expressions of the
representative accretion disk radii calculated in section
\ref{sec:inc} we refer the reader to the reviews by
\citet{1995CAS....28.....W} and \citet{2002apa..book.....F}. 

\subsection{Projected rotational velocity and binary mass ratio}\label{sec:q}

The motivation for the spectroscopic observations was to firmly
determine the binary mass ratio in J1820. To do this, we take advantage
of the fact that the tidal interactions between the BH and mass-donor star
are able of aligning orbital and  stellar spin axes, circularize
  the orbit, and  synchronize the stellar rotation with the  orbital
  motion. These three changes take place on timescales much shorter than the lifetime of the binary (see e.g. \citealt{1977A&A....57..383Z,1988ApJ...324L..71T}). Thus,  
the rotational broadening of the star, projected onto our line-of-sight, is %
$v_{rot} \sin i \approx K{_2}(1+q) R{_2}/a$,  where $i$ is
the binary inclination, $K_2$ the radial velocity semi-amplitude of
the donor star and $R{_2}/a$ the ratio between the Roche-lobe
effective radius of the donor star and the separation between the stellar
components. $R{_2}/a$ can be expressed  
as a function of $q$ only \citep{1983ApJ...268..368E} 
and, therefore,  the projected rotational velocity of the donor star 
is approximated to $v_{rot}\sin i \approx 0.49(1+q)q^{2/3}K_{2}[0.6
  q^{2/3}+\ln (1+q^{1/3})]^{-1}$. Hence $q$ can be  simply derived by measuring $K_2$ and $v_{rot} \sin
i$. The latter quantity is imprinted in the spectra by Doppler broadening of the donor star photospheric features.

To measure $v_{rot} \sin i$ we employ the optimal subtraction
technique described in \citet[see also
\citealt{2007ApJ...669L..85S}]{1994MNRAS.266..137M}. 
This technique relies on finding the smallest residual  
after subtracting a set of spectral templates (artificially broadened and scaled  to
match the width and depth of the  photospheric  lines in the donor star) 
from the Doppler-corrected average spectrum of the target. %
Our templates are the spectra of the three K-type dwarfs observed with
OSIRIS (Section \ref{sec:obs}), which have $v_{rot} \sin i \lesssim 2$ km s$^{-1}$ and spectral types close to that estimated for
the donor in J1820.  

Following \citet{2019ApJ...882L..21T}, we perform
the analysis in the spectral range $5200-6815$ \AA~excluding regions
containing emission lines, telluric or interstellar absorptions. The data were rebinned
onto a logarithmic wavelength scale and their continua normalized to
unity. The latter step was performed  by dividing each spectrum by  a
third-order spline fit  to the continuum. Radial velocity shifts between the nine normalized J1820 spectra and each spectral template were calculated by cross-correlation of features in the above
wavelength interval. Each J1820 spectrum was then
Doppler-corrected to the rest frame of  the template star by
subtracting the corresponding  radial velocity. The Doppler-corrected
spectra were averaged assigning different weights to the individual
spectra to  maximize the signal-to-noise ratio in the resulting sum.  
In order to account for  the 
radial velocity drift experienced by the stellar features 
during the exposures, due to the orbital motion of the donor star,  we
simulated the effect by making nine copies of each template,  which
were smeared considering the length of the exposures and the orbital ephemeris (see e.g. \citealt{2002MNRAS.334..233T} for details).  
The smeared template copies were then summed
using identical weights as for the J1820 spectra. 

The three summed
spectral templates were broadened from $1$ km s$^{-1}$ to $150$ km
s$^{-1}$ in steps of $1$ km s$^{-1}$ through convolution with the rotational profile of
\citet{1992oasp.book.....G}. We used limb-darkening coefficients of
$0.67, 0.69$ and $0.71$ for the K3 V, K4 V and K5 V templates,
respectively \citep{1995A&AS..114..247C} - use of a lower
limb-darkening coefficient (of 0.5) decreases   the $v_{rot} \sin i$
values reported in this work by $<3$ km s$^{-1}$. The gravity
  darkening and the Roche-lobe geometry of the donor star are not
  accounted for since the statistical uncertainties are dominant over
  the effect they introduce  in the  measurement of $q$. At this
  respect, see section 3.2 in \citealt{1994MNRAS.266..137M} for a thorough investigation of  their impact on  the evaluation of $q$ for the BH transient
  A0620-00. 

\begin{figure*}
\includegraphics[scale=0.59,angle=0]{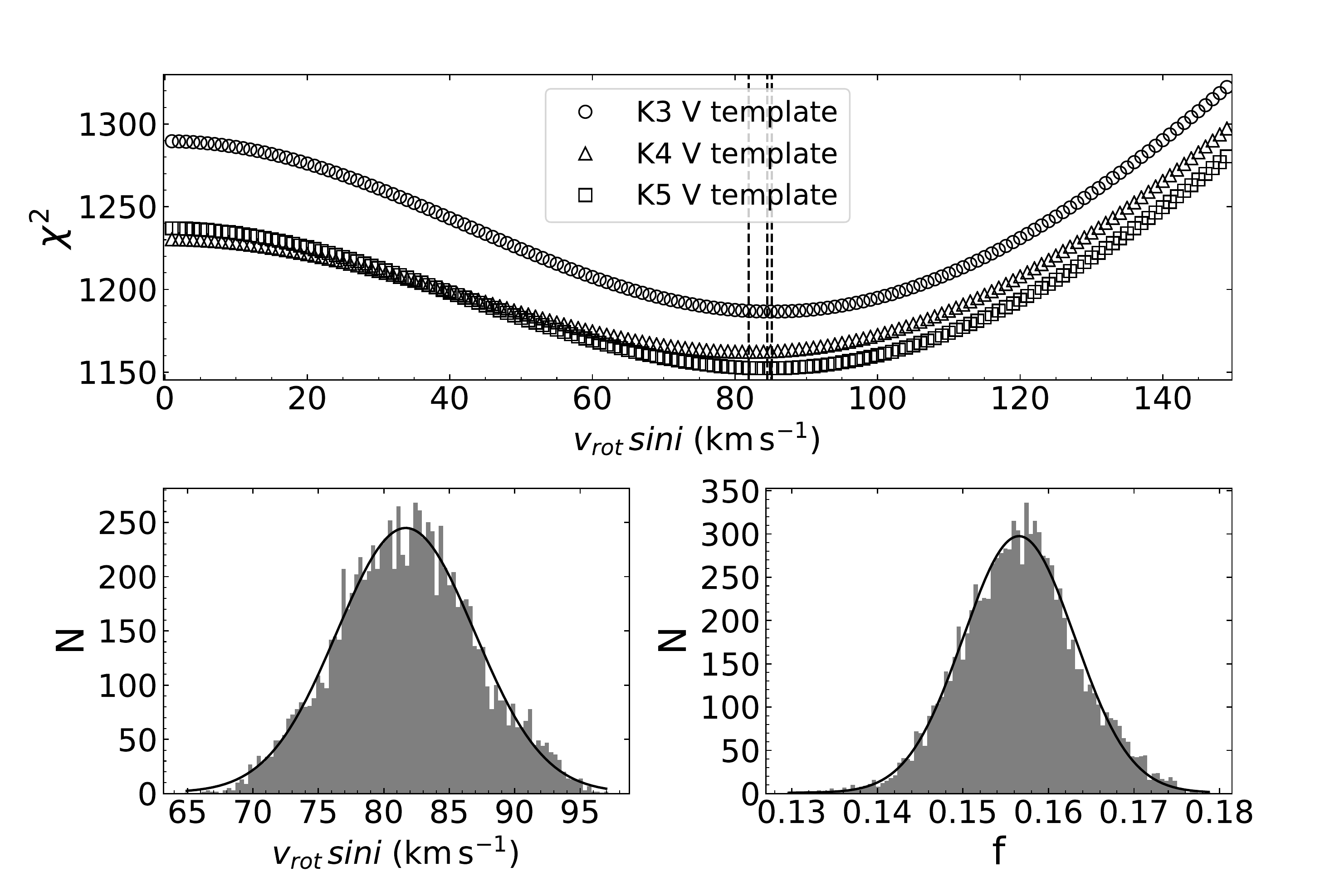}
\caption{Top: $\chi^2$ statistics as a function of  the $v_{rot}  \sin i$ applied 
to the K-type dwarf templates for its
optimal subtraction from the average J1820 spectrum (degrees of
freedom = 997). Bottom: distribution of
$v_{rot} \sin i$  values (left) and corresponding $f$ values (right)
obtained by subtracting the K4 V template from 10,000 bootstrap copies
of the J1820 average spectrum. Gaussian fits to the distributions are
shown. They deliver $v_{rot} \sin i = 82 \pm 5$ km s$^{-1}$  and $f = 0.158 \pm  0.007$
at the 68\% confidence level.}
\label{fig1}
\end{figure*}

The broadened templates were
subtracted from the  J1820 average spectrum while optimizing for a
variable scaling factor $f$ that represents %
the fractional contribution of light from the donor star to the total light in the binary system.  The final values for $v_{rot} \sin i$ and $f$ in J1820 were established  by
minimizing the $\chi^2$ statistics between the residual of the subtraction and a
smoothed version of itself. Figure \ref{fig1} shows for each template the
resulting $\chi^2$ as function of the applied broadening. The three
$\chi^2$ curves give single minima that yield  
consistent values of the optimal $v_{rot} \sin i$. To evaluate the  uncertainties, we
repeat the above optimal subtraction procedure and $\chi^2$ minimization in 10,000 bootstrap copies of the J1820 average
spectrum. All resulting distributions of $v_{rot} \sin i$ and $f$ were
satisfactorily described with a Gaussian model and we adopt their mean and
standard deviation as reliable estimates of their value and 1-$\sigma$
uncertainty. The results for each template are given in Table
\ref{vsini}.  The $\chi^2$ minimization  also shows that the  donor
star in J1820 contributes $\sim$16-17 \% to the total flux. This is accordant
with the contribution  found in the  data taken in June 2019 \citep{2019ApJ...882L..21T}. 

From the above results, we take $v_{rot} \sin i = 84$ km s$^{-1}$ and a 1-$\sigma$
uncertainty of 5 km s$^{-1}$ which, combined with $K{_2} = 417.7  \pm
3.9$ km s$^{-1}$, yields $q = 0.072  \pm 0.012$ (1-$\sigma$). We note here that a similar result ($v_{rot} \sin i = 81 \pm 8$ km s$^{-1}$) 
is achieved when applying this analysis to the higher resolution ISIS spectra in \citet{{2019ApJ...882L..21T}}, despite being taken under variable seeing-limited conditions.

\begin{table}
\begin{center}
\label{vsinil}
\caption{Optimal subtraction results (1-$\sigma$ errors).}
\begin{tabular}{lccc}
\tableline\tableline
Template     &Spectral        & $v_{rot} \sin i$ & $f$       \\
 	            &Type	          & (km~s$^{-1}$)  & \\ 
\\

HD 222237 &   K3~V   & $85 \pm 5$   & $0.173 \pm 0.008$   \\ 
HD 216803 &    K4~V  & $82 \pm 5$  & $0.158 \pm  0.007$  \\ 
61 Cyg A    &    K5~V  & $85 \pm 5$  & $0.17 \pm 0.01$\\ 
\tableline
\end{tabular}
\begin{tabular}{c}
\\
\end{tabular}
\end{center}
\label{vsini}
\end{table}     

\subsection{Outer disk radius and binary inclination}\label{sec:inc}

The analysis above has supplied us with a solid
measurement of $q$ that, in combination with information on the outer
disk radius, can serve to improve on the estimates 
of  the binary inclination. 

On the basis of geometrical arguments,  the non detection of X-ray
eclipses during outburst implies an upper limit to the binary
inclination of $\cos i \ge R_2/a = 0.49q^{2/3}[0.6q^{2/3}+\ln (1+q^{1/3})]^{-1}$. This condition yields $i < 80.8^{\circ}$ (3-$\sigma $). On the other hand, the detection of
a grazing eclipse of the disk by the donor star sets a lower limit to
the inclination of $\cos i \le R_2/(a- R_{\rm d})$ with $R_{\rm d}$
being the outer disk radius. This condition can be conveniently
rewritten as   
\begin{equation}
\cos i\le \frac{0.49q^\frac{2}{3}\left[0.6 q^\frac{2}{3}+\ln \left(1+q^\frac{1}{3}\right)\right]^{-1}}
{1 - \left[\frac{1}{2}-0.227 \log_{10}\left( q+0.01 \right) \right]
  \left(\frac{R_{d}}{b_{1}}\right) }
\label{eq:eclipse}
\end{equation}

\noindent by giving $R_{\rm d}$ relative to the inner Lagrangian point distance
to the BH ($b_1 $) and ${b_1}/a$ in function of $q$, following the
approximation in 
equation A2 of %
\citet{2018MNRAS.480.1580Z}. Thus, knowledge of $q$
and $R_{\rm d}/b_1$ constrains $i$. Estimates of $R_{\rm d}$ exist for three BH X-ray
transients in quiescence. These were established from detection in
H$\alpha$ Doppler tomograms of a well resolved and confined hotspot caused by the impact of the gas stream (from the donor star) with the outer disk
regions. These tomograms show that $R_{\rm  d}/b_1$ is
$0.50 \pm 0.05$ both in A 0620-00 ($q=0.067 \pm 0.01$;
\citealt{1994MNRAS.266..137M}) and Nova Muscae 1991 ($q=0.079 \pm
0.007$; \citealt{2015MNRAS.449.1584P}) while $R_{\rm  d}/b_1 =
0.47 \pm 0.03$ in  GS 2000+25 ($q=0.042 \pm 0.012$;
\citealt{1995MNRAS.277L..45C}). 
An H$\alpha$  tomogram  performed for J1820 by phase folding the
data obtained in June-August 2019 on the orbital period did not lead
to the detection of a  hotspot.  Based on the tomograms of the above three quiescent BH binaries, we adopt $R_{\rm  d}^{\rm quie}/b_1 \sim 0.5$
for J1820 in quiescence, a value that is near the stream circularization radius $R_{\rm
  circ}/b_1 = (1+q) (b_1/a)^3 = 0.45$. Presumably,  at the time of
the spectroscopic detection of the grazing eclipse in J1820, the disk
was larger than in  true quiescence since the system was still fading
in brightness. This can be tested by using the FWHM of the H$\alpha$
emission line as proxy of the disk size. Thus, we 
calculate an expected H$\alpha$ FWHM$^{\rm quie}$ of $1793 \pm 101$
km s$^{-1}$ for J1820 in true quiescence using the empirical relation
$K{_2} = (0.233\pm 0.013) \times {\rm{FWHM}}^{\rm quie}$ found for
quiescent BH transients \citep{2015ApJ...808...80C}. Following the procedures outlined by this author, we measure
the H$\alpha$ FWHM in the spectroscopy presented in
\citet{2019ApJ...882L..21T}.  We derive a FWHM of 
 1614 km s$^{-1}$  and standard deviation 96 km s$^{-1}$  by fitting a single
Gaussian model to the individual normalized line profiles while
correcting for  the instrumental broadening. Assuming a $r^{-1/2}$
rotational (Keplerian) velocity law for the disk, we obtain $R_{\rm
  d}/R_{\rm d}^{\rm quie}= ({\rm{FWHM}}^{\rm quie}/{\rm{FWHM}})^2= 1.2
\pm 0.2$. This implies  $R_{\rm  d} \sim  (0.60  \pm 0.10) b_1 =
(0.74 \pm 0.12) R_1$ at the time of the grazing eclipse detection. 
Here $R_1$ is the equivalent radius of the  compact object's Roche lobe.
Admittedly, this calculation does
not allow us to set a stringent value for $R_{\rm  d}$ and we are left
with imposing limits to  this quantity and $i$ by employing
characteristic accretion disk radii. A lower limit to $i$ comes from  the
maximum reachable disc radius caused by tidal truncation: $R_{\rm
  t} = 0.6/(1+q) a = 0 .75 b_1 = 0.93 R_1$. On the other hand, during
outburst  the disk had to expand to or beyond the
3:1 resonance radius 
(or radius of the $j=3, k=2$ commensurability) %
$R_{32} = 0.63 b_1 = 0.78 R_1$, a condition that drives the superhump
phenomena (e.g. \citealt{1991MNRAS.249...25W}).  Given that these analytically inferred
radii are solely dependent on $q$, the evaluation of equation
\ref{eq:eclipse} was performed through Monte Carlo ramdomization with
$q$ treated as being normally distributed. This yielded  normalized
distributions for $i$ from which we obtain 3-$\sigma$ lower limits of
$61.9^\circ$, $66.2^\circ$ and $72.8^\circ$ for a disk  radius equal
to $R_{\rm  t}$, $R_{32}$ and $R_{\rm  circ}$, respectively. Therefore, 
from geometrical arguments alone we conservatively constrain $i$ to be in the range $61.9{^\circ}-80.8{^\circ}$. 
 
\section{Discussion}\label{sec:disc}

\begin{figure*}
\includegraphics[scale=0.60,angle=0]{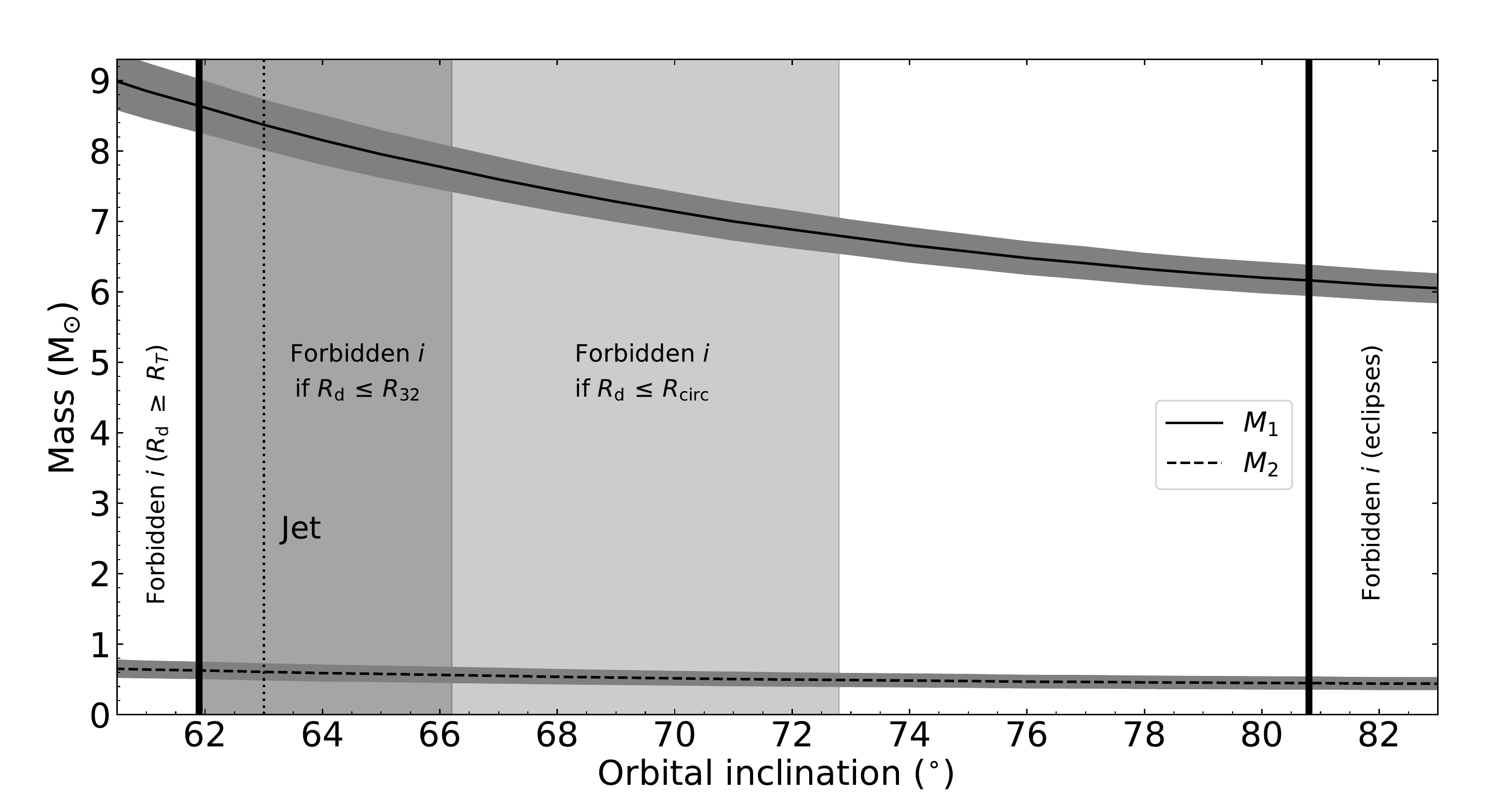}
\caption{Possible values for the BH and donor mass ($M_1$ and
  $M_2$, respectively) as a function of the orbital inclination. 
The encompassing shaded regions show their 1-$\sigma$ uncertainties.
The solid vertical lines mark extreme limits on the inclination: $i$ must be
$<80.8^{\circ}$ since the system lacks X-ray eclipses,
while $i>61.9^\circ$ is imposed by the condition that the disk size
near quiescence shall not pass its tidal truncation radius, $R_{\rm  T}$.
The dark grey-shaded area marks impossible mass solutions if the
outer disk radius $R_{\rm  d}$  reached or exceeded the 
 3:1 resonance radius $R_{32}$ at the time of the grazing eclipse detection with a
minimum possible value for $i$ of $66.2^\circ$.  The light grey-shaded
area represents impossible solutions if the accretion disk shrank to its
circularization radius $R_{\rm  circ}$. In this case, the minimum
possible $i$ is $72.8^\circ$. A vertical dashed line marks the
inclination found for the radio jet.}
§§\label{fig2}
\end{figure*}

The analysis of the OSIRIS spectroscopy (section \ref{sec:q}) has yielded
$q = 0.072 \pm 0.012$, which is a factor $\sim2$ lower than  the value
estimated in our previous work by employing  the relation between this binary
parameter and the fractional superhump period
excess $\Delta P = (P_{sh} - P_{orb})/ P_{orb}$ \citep
{2019ApJ...882L..21T}. This discrepancy is most likely
due to the use in the calculation of $\Delta P$ of  the 
preliminary $0.703 \pm 0.003$ d superhump period ($P_{sh}$) provided in \citet
{2018ATel11756....1P}. An updated (but still provisional) period of
$0.6903 \pm 0.0003$ d for the superhump was 
later reported by \citet{2019P} from further time-series photometry,
while evidence for early evolution of $P_{sh}$ was communicated by Variable Star Network observers that
followed the 2018 outburst. Thus, a valid comparison between the
result presented in this work and that obtainable through $\Delta P$
is subject to the publication of a rigorous study of the superhump
temporal properties in J1820. 

In section \ref{sec:inc} we constrain the binary
inclination to be in the range $61.9{^\circ} < i < 80.8{^\circ}$. These
limits supersede the early restriction of $69{^\circ} \lesssim i
\lesssim 77{^\circ}$ calculated by  adopting an outer disk
radius of $R_{\rm d} \sim 0.5 R_1$ and $q\simeq 0.12$
\citep{2019ApJ...882L..21T}.  On one hand, the allowed range for $i$
encloses those expected for an X-ray dipper binary
\citep{2002apa..book.....F} and contains the  $63 \pm 3^\circ$
(1-$\sigma$) inclination found for the radio jet
\citep{2020MNRAS.493L..81A}. On the other hand,
it excludes the $29.8^{+3.0}_{-2.7}{^\circ}$ (3-$\sigma$) inclination for the inner disk
obtained from modeling of X-ray spectra \citep{2019MNRAS.487.5946B}. Still, we consider the new  lower boundary of $61.9{^\circ}$
imposed by the disk tidal truncation radius  not likely because the
accretion disk should have contracted appreciably during the approach to
quiescence.  In fact, the H$\alpha$ FWHM at the time of the grazing
eclipse delivered a mean value of $R_{\rm  d} \sim 0.6 b_1$ (section \ref{sec:inc}) which is
close to the 3:1  
resonance radius ($R_{32} = 0.63 b_1$). Thus, the
grazing eclipse occurrence and standard deviation associated with the value of
$R_{\rm  d}$ evaluated from H$\alpha$ can potentially be explained as
caused by an asymmetric  accretion disk 
extending to the resonance radius. In such a setting, the range of allowed inclinations
narrows to $66{^\circ} < i < 81{^\circ}$. Support  for this
scenario could be provided by  the photometric or spectroscopic
detection of a precessing  disk at the time of our observations or
during early quiescence of the source (e.g. \citealt{2002ApJ...569..423T,2002MNRAS.333..791Z}).

Knowledge of $K_{2}$, the orbital period and $q$ allows us to give
 the BH and donor star masses as a function  of the binary inclination

\begin{align*}
M_1 &= \frac{5.95\pm0.22}{\sin ^3 i} M_\sun, & M_2 &= \frac{0.43\pm0.08}{\sin ^3 i} M_\sun  
\end{align*}
\newline
\noindent We display in Figure \ref{fig2} the expected values of $M_1$ and
$M_2$ obtained from  both expressions for a given $i$ with
assumed 1-$\sigma $ uncertainty of $1^\circ$. Taking the favoured
constraint on the inclination ($66.2{^\circ} < i < 80.8{^\circ}$) and
the 1-$\sigma$ uncertainties in the above expressions, we derive  the
following ranges for the masses at $1 \! - \! \sigma$:

\begin{align*}
 5.96 &< M_1(M_\odot) < 8.06, &  0.36 < M_2(M_\odot) < 0.66  
\end{align*}
and $2 \! - \! \sigma$:
\begin{align*}
 5.73 &< M_1(M_\odot) < 8.34, &  0.28 < M_2(M_\odot) < 0.77 
\end{align*}

The dipping behaviour in J1820 was not as complex and copious as observed during the 
outburst of the 0.406 d orbital period BH candidate MAXI J1305-704
\citep{2013ApJ...779...26S}, which has a close matching orbital period
among known BH X-ray dippers. We take this difference as a supporting argument to
prefer  the low-end values in the range of possible binary inclinations
found for J1820. This, in turn, points to large values of the
permitted  BH mass and to the
less undermassive (evolved) donor star.
It should be noted that the latter has spectral type K3-K5 and main-sequence stars with this spectral type have masses of  $\sim0.7 M_\odot$. %

We must stress here that the limits on $i$ calculated in this work rest in the use of a
simple model for the binary geometry that ignores the disk vertical
structure and therefore they are subject to systematic effects. In
this respect, the opening angle subtended by the accretion disk has been
constrained observationally to $\leq20^\circ$ in neutron star low-mass X-ray binaries
(see e.g. \citealt{2018MNRAS.474.4717J} and references there
in). Hence, if we assume a similar angle for the disk in quiescent BH transients,
the lower boundary on $i$ will decrease such that the $63 \pm
3^\circ$ (1-$\sigma$) orientation angle for the jet ejecta
\citep{2020MNRAS.493L..81A} is fully compatible with being the binary
inclination. If that was the case, we obtain values for the
  BH and donor star masses (at the 68\% confidence level) of 
\begin{align*}
M_1 &= 8.48^{+0.79}_{-0.72} M_\odot, & M_2 &= 0.61^{+0.13}_{-0.12} M_\odot 
\end{align*}

\noindent Modelling of the ellipsoidal
modulation of the donor star expected in lightcurves obtained in
the quiescence state may refute this possibility. 

\section{acknowledgements}
We are thankfull to the GTC staff, in particular Antonio
L. Cabrera Lavers, for their help to implement the
spectroscopy presented in this paper.  
We thank the referee for the useful comment.
We acknowledge support by the Spanish MINECO under grant
AYA2017-83216-P. TMD and MAPT acknowledge support via Ram\'on y Cajal
Fellowships RYC-2015-18148 and RYC-2015-17854.  
PGJ acknowledges funding from the European Research Council under ERC
Consolidator Grant agreement no 647208. MH acknowledges an ESO fellowship.
{\sc molly} software by Tom Marsh is gratefully acknowledged. {\sc
  iraf} is distributed by the National Optical Astronomy Observatory,
which is operated by the Association of Universities for Research in
Astronomy (AURA) under a cooperative agreement with the National Science Foundation.



\end{document}